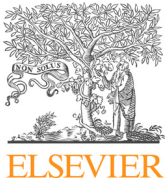

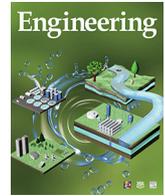

Research
Managing Projects under Uncertainty—Review

# Stochastic Earned Duration Analysis for Project Schedule Management

Fernando Acebes *, David Poza, José Manuel González-Varona, Adolfo López-Paredes

*GIR INSISOC-Department of Business Organization and CIM, School of Industrial Engineering, University of Valladolid, Valladolid 47011, Spain*

ABSTRACT

Earned duration management (EDM) is a methodology for project schedule management (PSM) that can be considered an alternative to earned value management (EVM). EDM provides an estimation of deviations in schedule and a final project duration estimation. There is a key difference between EDM and EVM: In EDM, the value of activities is expressed as work periods; whereas in EVM, value is expressed in terms of cost. In this paper, we present how EDM can be applied to monitor and control stochastic projects. To explain the methodology, we use a real case study with a project that presents a high level of uncertainty and activities with random durations. We analyze the usability of this approach according to the activities network topology and compare the EVM and earned schedule methodology (ESM) for PSM.

© 2021 THE AUTHORS. Published by Elsevier LTD on behalf of Chinese Academy of Engineering and Higher Education Press Limited Company. This is an open access article under the CC BY-NC-ND license (http://creativecommons.org/licenses/by-nc-nd/4.0/).

## 1. Introduction

Monitoring and control activities are a crucial aspect of project management throughout a project's life cycle [1,2]. Earned value management (EVM) has been the most widely used tool for project control since 1967 [3]. The US Department of Defense (DoD) approved a directive that includes key EVM parameters among the 35 criteria that should be met by any industrial firm when applying some kind of cost-reimbursable or incentive contract for major system procurements. The US DoD provided an EVM system (EVMS), which was initially and inflexibly applied for decades by the US Government; later, it was also adopted in other countries such as Australia, Canada, and Sweden. The key point in adopting EVM was to prevent cost growth risk when the government is the final responsible party for budget overruns. EVM was initially adopted for monitoring and controlling project costs.

A detailed explanation of EVM can be found in Refs. [3–5] as well as in the extensive list of references [6–8]. Lipke [9] proposed the earned schedule (ES) with EVM parameters to provide new metrics for schedule control. Pajares and López-Paredes [10] considered the inherent uncertainty of activities and defined the schedule control index (SCoI) and cost control index (CCoI) indicators to overcome some limitations in ES and EVM for schedule control. Acebes et al. [11] went one step further and described a graphical framework for integrating cost, schedule, and risk monitoring.

Khamooshi and Golafshani [12] defined an alternative framework to EVM, known as the earned duration management (EDM) methodology. EDM is intended for monitoring and controlling the project schedule by redefining EVM parameters in work-period terms rather than cost measurements [13]. Comparisons between EVM and EDM frameworks are a fruitful field for researchers interested in project control and the accurate forecasting of final costs and duration. De Andrade and Vanhoucke [14] presented the results they obtained from comparing EDM and EVM in the provision of accurate project duration forecasting. Using data from real projects, they concluded that EDM provided better results when using the EDM project regularity indicator.

Our approach to improve the accuracy of both project control and final duration forecasting is to consider inherent aleatoric uncertainty when estimating the work periods for each activity in the planning stage. The starting point is the process described by Acebes et al. [15] to improve project monitoring and control within the EVM framework. We adapted this process to the EDM in order to acquire better knowledge of any deviations from the planned schedule at any intermediate control milestone, as well as the final project duration estimation. With the results obtained in this way, and by taking data from real projects (with significant differences in network topology), we compared EVM and ES using the mean absolute percentage error (MAPE).

This paper is structured as follows. In Section 2, we present a bibliographic review of the most relevant contributions for project monitoring and control by the EVM methodology. We then present the research process followed in this paper. We describe stochastic earned duration analysis and project control techniques and offer a case study to better illustrate the process of applying the methodology to a real project. Next, we compare the use of EVM and earned schedule methodology (ESM) for projects with different network topologies. Finally, we present the main conclusions and results.

---

* Corresponding author.
  *E-mail address:* fernando.acebes@uva.es (F. Acebes).

## 2. Background

Project monitoring and control processes are transcendental functions in project management [1,16]. They are designed to take information from the status project execution and compare it with the base scenario. Analyzing possible planning variance can be useful in the decision-making process of adopting potential corrective measures [17,18].

This section provides a general summary of the research conducted on project control methods and on the different statistical algorithms that we use in our proposed methodology. First, we discuss project control using the EVM methodology. Second, we examine project control using the EDM. In the last section, we describe the different statistical techniques that are applied to the classification and regression problems.

### 2.1. Progress monitoring: EVM

Rozenes et al. [19] conducted a literature review to provide an overview of the nature and importance of project control, including factors that determine the project success and control systems analysis. The most recent literature focuses mainly on EVM as the most widely used fundamental tool for project control, specifically when monitoring cost [20] and controlling time based on the ES concept [9].

Willems and Vanhoucke [21] classified the bibliography that relates project control to EVM methodology. One of the classifications conducted in their work was based on the methodology that has been applied to problem solving. Another classification that Willems and Vanhoucke [21] included in their work was established in accordance with the degree of uncertainty inherent to each project: deterministic, stochastic, and fuzzy. Thus, there are analyses whose calculations are based on average or expected values that are labeled as deterministic. Some analyses consider the uncertainty of project activities as probability distribution functions, which are labeled as stochastic. The result consists of distributions and confidence intervals connected to estimates. Finally, there are fuzzy analyses that take a probabilistic approach in which data are not only imprecise, but also vague. Hence, they can be represented by fuzzy numbers and manipulated by fuzzy techniques [22].

The literature on deterministic and stochastic analyses is very extensive. Pellerin and Perrier [1] compiled works dealing with methods, techniques, and tools for project planning and control by paying special attention to the EVM in the project control section. Hazır [18] conducted a similar study by identifying analytical models and decision support tools for project control. He included the work by Pajares and López-Paredes [10], who introduced a new procedure for project control in an environment with uncertainty (the procedure extended in Ref. [11]). Colin and Vanhoucke [23] focused their study on process control during project execution and compared the efficiency of different control methods using EVM. Vanhoucke [2] also reviewed works on tolerance limits in process control. All the articles cited here include many others that have been reviewed, compared, and analyzed, providing us with a measure of the importance attached to the EVM in relation to project control [6,23–31].

As for research about the estimation of project duration upon its completion, an extensive bibliography exists [32–34]. Batselier and Vanhoucke [33] evaluated the accuracy of different forecast techniques based on EVM. Wauters and Vanhoucke [35] focused on the stability of the results offered by EVM in comparison with computational experiments. Batselier and Vanhoucke [36] assessed different prediction methods using EVM applied to real projects to compare the obtained results. Finally, although the compilation can be widely extended, we mention the work by Wauters and Vanhoucke [37], who reviewed different methods that incorporate artificial intelligence for forecasting the final project duration.

Both fuzzy techniques and EVM have been widely used by several authors. Naeni et al. [38] present a fuzzy-based earned value model, in which they incorporate uncertainty from people's judgements and transform it into linguistic terms. Mortaji et al. [39] employed EVM in vagueness environments using left–right (L–R) fuzzy numbers. Salari et al. [40] used fuzzy techniques with EVM in financial aspects of the cost control system. Salari et al. [41] applied EVM to predict future project performance through statistical modeling. Due to the vagueness and imprecision associated with the data from real case projects, the time and cost behavior of each option were presumed to be fuzzy numbers. Similarly, other authors have used fuzzy techniques with EVM to improve a project's future performance by forecasting [41–44].

It should be noted that the fuzzy technique is based on the use of linguistic terms when assigning a specific value to a variable is impossible. Above all, such variables could be considered ambiguous, imprecise, or vague. Something similar occurs with the grey system theory [45], which can be applied together with the EVM [46]. Although these two methodologies (fuzzy and grey system techniques) work with uncertainty, they can be differentiated from stochastic analyses, in which project activities are perfectly defined for their probability distribution functions.

With time, continuous innovations have been proposed in the methodology that have attempted to improve both future predictions and project control at each monitoring time. All these refinements to the methodology, in both research branches, are part of attempts to obtain more effective results with fewer errors (in both project control and forecast).

### 2.2. Progress monitoring: EDM

Khamooshi and Golafshani [12] introduced the EDM concept. To remove the correlation between project cost and duration in the EVM methodology, they proposed a methodology that measures the "work" carried out during each period. The control of project duration, as well as the forecast of its final duration at each control time, are accomplished in duration terms independently of the cost of each activity.

Although the ES technique proposed in Ref. [9] uses time units for the control of the project timeframe, the calculation of this parameter (i.e., ES) is based on cost units. By using EDM, the calculations for project duration and cost become independent, and each employs the units corresponding to its magnitude. Since the novel EDM was first introduced, several studies about it have appeared. De Andrade and Vanhoucke [14] performed a comparative analysis between time predictions using ES and EDM. Khamooshi and Abdi [47] applied the EDM to predict project duration with exponential smoothing techniques. Ghanbari et al. [48,49] applied fuzzy techniques to solve uncertainty problems by applying the EDM. The literature involves the use of fuzzy techniques in conjunction with EDM [48,50–52].

De Andrade et al. [13] used real projects to compare project duration estimation results according to both EDM and EVM.

Yousefi et al. [53] controlled the project by applying statistical control charts based on the indices provided by the EDM methodology.

Since the concepts and indicators provided by the EDM were first introduced, we have observed that many innovations are related to them; thus, we aim to use this methodology to improve project monitoring and control, as well as forecasts of duration and final cost. In this paper, we present a methodology based on EDM regarding project control that makes predictions of a project's final duration by incorporating uncertainty into project activities through a probability distribution function.

### 2.3. Algorithms for classification and regression problems

This section explains the basic concepts of the statistical analysis used herein, as we briefly indicate what we mean by anomaly detection, classification, and regression. We also briefly describe the different algorithms used for these techniques.

Novelty detection consists of identifying observations that derive from, or are inconsistent with, the sample data in which they occur [54,55]. The idea is to build a model that describes the normal project behavior range. This normality model is used as a test by comparing it with the actual project development.

The methodologies used for novelty detection focus on estimating the generative probability density function from the data drawn from the simulation data. This function is used to calculate the probability of a new observation (the actual project) having been generated by the distribution [54–57]. We applied the kde2d function of the "MASS" package of R software for kernel density estimations with radial kernels [58].

Analyzing the data as a classification problem allows us to estimate the probability of a project finishing on time. A classification problem aims to predict a quantitative variable, which is often referred to as a response, outcome, or dependent variable with a set of qualitative and/or quantitative variables called predictors, independent variables, or simply variables.

Analyzing the data as a regression problem allows us to quantify any project lag. A regression problem involves predicting a qualitative, qualitative, or continuous variable, also called a response, output, or independent variable, as in the classification problem, with a set of qualitative and/or quantitative variables, the predictors.

We now go on to explain the operation of all the algorithms that we use in our model, which are all included in the R software package "caret" [59,60]. The linear discriminant analysis (LDA) is a dimensionality reduction technique. It is used as a pre-processing step in machine learning and pattern classification applications. The goal of LDA is to project features in a higher dimensional space onto a lower-dimensional space to avoid the dimensionality curse and to reduce the required resources and dimensional costs [61,62].

A classification and regression tree (CART) is a predictive model that explains how an outcome variable's values can be predicted based on other values. CART output is a decision tree in which each fork is split into a predictor variable and each end node contains a prediction for the outcome variable [63]. K-nearest-neighbor (kNN) is a supervised instance-based machine learning algorithm. It can be used to classify new samples (discrete values) or to predict (regression and continuous values). It essentially classifies values by finding the "most similar" (in terms of closeness) data points learned in the training stage and by making assumptions about the new points based on that classification [64].

Support vector machines (SVMs) are a set of supervised learning algorithms. These methods are usually related to classification and regression problems. With a set of training examples (of samples), we can label classes and train an SVM to build a model that predicts the class of a new sample. Intuitively, the SVM is a model that represents the sample points in space by separating classes into two spaces that are as wide as possible by defining a separating hyperplane as the vector between the two points of the two classes that are closer to one another, known as the support vector. When the new samples are put in correspondence with this model, depending on the spaces in which they belong, they can be classified as one class or the other [65]. A random forest (RF) is a combination of predictor trees, where each tree depends on the values of an independently tested random vector with the same distribution for each tree. It is a substantial bagging modification that builds a large collection of uncorrelated trees and then averages them [66]. Linear regression (LR) is a linear approach that is followed to model the relation between a scalar response and one explanatory variable or more (also known as dependent and independent variables). The case of one explanatory variable is called simple LR; the process is called multiple LR if there is more than one [67].

## 3. Stochastic earned duration methodology (SEDM)

Acebes et al. [15] describe a methodology for project monitoring and project control called stochastic earned value analysis. Its starting point is to generate a wide range of projects by means of a Monte Carlo simulation that are compatible with the planned project's specifications. The analysis at each control milestone depends on the statistical techniques that are followed to study the project—that is, anomaly detection algorithms, and classification and regression problems[†]. The ultimate aim is to provide project managers with a decision support system to detect abnormal deviations from the planned project and to estimate the probability of overruns, as well as the expected time and work-periods until the project ends.

### 3.1. Earned duration management

EDM is a methodology that creates duration-based performance metrics and aims to decouple schedule and cost performance measures completely. EDM focuses on the exclusive usage of time-based data for the generation of physical progress indicators. Rather than considering the value of project activities in monetary units, the value of activities is expressed as work periods. This is the key difference among EDM, EVM, and ES, as the latter two are based on the costs of activities.

With this information, in the planning phase, we can create the curves displayed in Fig. 1: the total planned duration (TPD), total earned duration (TED), and total actual duration (TAD) [12]. The analogy with those used in the EVM is evident: ① TPD is the cumulative number of planned working periods throughout the project; ② TAD comprises all the working periods spent prior and up to the actual duration (AD); ③ TED is the number of working periods earned by AD (i.e., the value of the performed work (e.g., workdays) expressed as a proportion of the planned work).

As with the EVM, the TED value equals the final planned TPD value at the end of the project.

By definition, each planned day of every activity has a weight of 1, regardless of the effort, resources, or costs involved in performing it. The AD of each activity $i$ (AD$_i$) is the number of working days that it took to complete that activity. To calculate the value of the daily earned duration (ED) for every activity effectively performed on a working day, the planned duration (PD) of an activity $i$ (PD$_i$) must be divided by its AD$_i$. The sum of the daily EDs of a certain activity $i$ defines the ED of that activity (ED$_i$). The ED of a project calculated during a control period (TED) corresponds to the sum of the $ED_i$'s of all the project activities.

---

[†] This analysis can be extended by adding new algorithms (i.e., statistical or machine-learning techniques).

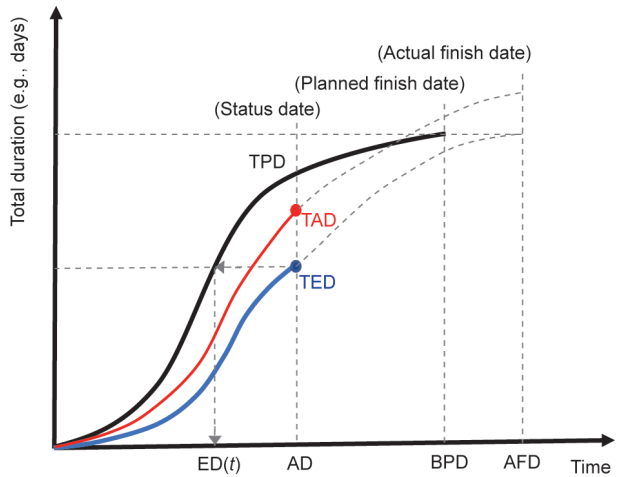

**Fig. 1.** Conceptual EDM graph based on Ref. [12]. ED($t$): the earned duration (ED) corresponding to the control period; AD: actual duration; BPD: baseline planned duration; AFD: actual finish date.

Fig. 1 represents a conceptual EDM graph based on Ref. [12]—an S-curve that depicts the total accumulated duration for the planned (and actual) progress of activities on each working day for a project plotted against time. The magnitude represented on the *y*-axis in this methodology is the accumulated sum of the planned time units that correspond to the activities performed during that period. Therefore, the graph takes on a theoretical "S" form, with a greater or lesser approximation to an "S" curve depending on each project's scheduling.

AD is the control milestone point that we wish to use to monitor the project. It is the equivalent to the actual time in EVM/ES. At this time, we know the TED, and we can calculate the ED($t$) on the TPD curve. Eq. (1) is the analytical expression used to calculate ED($t$):

$$ED(t) = AD + \frac{TED - TPD_t}{TPD_{t+1} - TPD_t} \tag{1}$$

where ED($t$) is the ED corresponding to the control period. Note that $t$ and $t + 1$ are two consecutive time periods that belong to the TPD curve such that $t < ED(t) < t + 1$ and $t + 1 - t = 1$ (if we consider calendar unitary).

The calculation of ED($t$) corresponds to the resolution of a linear interpolation, where the TPD curve approaches a straight line between time instants $t$ and $t + 1$ (which are known). This calculation is independent of the TPD curve shape if the curve shape takes an S-curve form or if the representation is linear.

### 3.2. Stochastic earned duration analysis

Our analysis is performed in two stages. First, in the planning phase, we collect the information available about the aleatoric uncertainty of the activities (i.e., the type of probability distribution function and the characteristic parameters such as the expected value, standard deviation, most optimistic end date, most likely end date, most pessimistic end date, etc.). We apply the Monte Carlo simulation to generate a large population of simulated projects. These "instances" of the approved planned project are the universe of simulated planned projects that are compatible with the available information (i.e., the network topology and the uncertainty of the activities). The type of distribution function assigned to each activity does not affect the representation of the indicators illustrated in Fig. 1 (TPD, TAD, and TED), as the expected values of the timework units of the activities are used to represent the indicators. The type of distribution function has an impact on the values assigned to activities when applying the Monte Carlo simulation. Therefore, this would impact the actual situation of the project that is underway in comparison with the planned project at each control point, as will be seen, for example, in Section 4.1 below.

In the second phase, we use advanced statistical techniques at each control milestone during the project. Our goal is to answer the questions that project managers ask: Do we have to take corrective action or can the observed deviations from the planned project be considered to be in accordance with the expected behavior? Can we obtain an accurate estimation of the final time when the project ends?

In the following subsections, we provide a detailed explanation of the methodology, along with a flow diagram of the whole process at the end of this section.

### 3.2.1. Planning phase: Monte Carlo simulation

In the project planning phase, we have information about the activities that make up the project—namely, the sequencing, duration, and aleatoric uncertainty that defines them. By using the most probable duration for the activities and applying EDM, we can represent the TPD curve, as shown in Fig. 2.

Because the duration of the project activities contains aleatoric uncertainty, we can apply the Monte Carlo simulation to the project under study. As a result, we obtain a large number *N* of possible simulated projects that are compatible with the uncertainty defined for each activity. Each of these simulated projects represents a possible project execution, because every activity was randomly assigned a duration following the distribution function with which it was programmed.

For each of these simulated projects *j*, we can construct its $TAD_j$ curve of real duration (RD); we can also calculate its corresponding $TED_j$ curve of ED. In Fig. 2, we display the planned project (TPD), the project underway at $t =$ AD, and only one simulated project instance (*j*).

As with the EVM, where, by definition, PV = EV at the end of the project (independently of the project being delayed or continuing), in the EDM at the end of the project, TED = TPD (TPD$_{BPD}$ = TED$_{AFD}$) (where PV is planned value; EV is earned value; BPD is baseline planned duration; and AFD means actual finish date). Therefore, all the $TED_j$ curves of each simulated project will take the same value at the end of the project, which will coincide with the TPD value at the end of the project (TPD$_{BPD}$ = TED$_{JAFD_j}$) (Fig. 2). During the proposed process, all the projects that we obtain after applying the Monte Carlo simulation will have the same planned project in common (the same TPD curve). Therefore, if TED$_{AD}$ corresponds to a percentage of the final value of TPD, the TED$_{AD_j}$ of any simulated project will coincide with the previous value because, if we apply the same percentage to the identical final value (TED$_j$ = TPD), the TED value of each curve will coincide (TED$_{AD}$ = TED$_{AD_j}$).

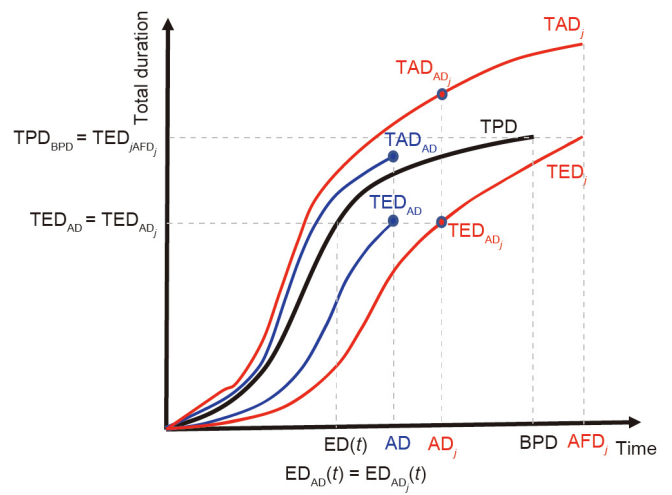

**Fig. 2.** Monte Carlo simulation: planned project (TPD), project underway (TED$_{AD}$ and TAD$_{AD}$), simulated project *j* (TED$_{AD_j}$ and TAD$_{AD_j}$), and the BPD.

At each control milestone AD, we know the pair (TED$_{AD}$, TAD$_{AD}$) for the project underway. There are *N* simulated projects, and we can calculate in AD$_j$ the triad (AD$_j$, TED$_{AD_j}$, TAD$_{AD_j}$) from TED$_{AD}$ for each simulated project. We calculate in AD the project progress index (PPI) for the project that is underway (Eq. (2)).

$$PPI = \frac{ED(t)}{BPD} \qquad (2)$$

where BPD is the planned finish date. ED($t$) is equal for the project underway and for every simulated project *j* because it is obtained during the project planned with TED$_{AD}$, which is the same for each simulated project TED$_{AD_j}$.

From the *N* simulated projects, we can obtain a point cloud of pairs (AD$_j$, TAD$_{AD_j}$) for this PPI value. We use these data to obtain the statistical properties for the project at any control milestone and to know if project deviations are a consequence of the expected variability. We then apply advanced statistical techniques to these data, which provide the benchmark for real project monitoring and control (classification problem), and for forecasting the expected duration (regression problem).

### 3.2.2. Control phase: Project control and predicting final duration

Following the analysis performed in Ref. [15], we apply the techniques developed for anomaly detection to study the project deviations. The ultimate purpose is to build a model that describes the project's "normal" range of behavior. We need to aggregate all the point clouds (AD$_j$, TAD$_{AD_j}$) obtained previously for each PPI value to gain a general density function (Fig. 3). We use function kde2d, which is included in R "MASS" [58]. At each control milestone of the real project that is underway (AD, TAD$_{AD}$), we can calculate the probability that falls within the expected variability (Fig. 4).

To estimate the probability of the project finishing on time, we analyze data as a classification problem; we use the data from the *N* simulated projects. The algorithms use a proportion of the data for model training, and the rest is used for the trial. In the control milestone AD, we know the values (AD$_j$, TAD$_{AD_j}$) for each simulated project and whether the project will finish on time or not. We use the R "caret" software package [59,60] first to cross-validate the data employed for the training and trial. Afterward, this package makes it possible to test different algorithms to select the one that performs better for the studied classification problem [78–82]†. We select LDA, CARTs, *k*NN, SVMs with a radial kernel, and an RF. Finally, we choose the algorithm that performs best in our classification problem. In Section 4, we show the process in a real case study.

To accurately forecast the time when the project that is underway is expected to finish, we must study the data from simulated projects as a regression problem, as we do in the classification problem. In this case, from each simulated project *j*, we use the absolute value of the deviation (advance or delay) of its final duration from the planned value (BPD). We also employ the R "caret" package to test the following algorithms: LR, generalized linear regression (GLM), penalized linear regression (glmnet), CART, SVMs with a radial basis function, and *k*NN. We must also choose the best algorithm for the regression problem, as explained in the next section.

To evaluate and compare the accuracy of the different estimation methodologies, we use the MAPE measure, which has been previously employed in related research [68–71]. This measure provides a percentage value of the forecasting method's predictive accuracy (Eq. (3)). The lower the MAPE value for a forecasting method is, the more accurate the method is.

$$MAPE = \frac{100\%}{n} \sum_{t=1}^{n} \frac{|RD - EDAC^t|}{RD} \qquad (3)$$

where *n* refers to the total number of monitoring periods while the project is underway. Estimated duration at completion (EDAC) is used to indicate the final estimated project duration, calculated while period *t* is monitored.

To calculate the MAPE, we use the following as data: RD and the forecast value (EDAC). The former is obtained from the prediction of our model during each control period (EDAC$_t$). The formula does not use the AD in the calculation, but the RD. At each control time, the error between the real project duration (RD) versus the forecast duration (EDAC$_t$) can be calculated.

## 4. Computational experiment

To explain the SEDM, we choose a real project from the OR-AS database [72–74]. We select the project called "2016–15 Residential House Structural Work.xlsx," which consists of 13 activities with a planned duration (BPD) of 126 time units. In the end, however, the project ended after a duration of 130 time units. For the work periods, 141 workdays were planned, but the real final work took 151 workdays.

BPD is the project planned duration. After calculating all the activity durations and sequencing them properly, we conclude that the total project planned duration in the example is 126 time units. In the EDM, we calculate the timework units for each activity. By definition, each planned timework unit (day, week, month, etc.) of each activity has a weight of 1, regardless of the effort, resources, or costs involved in its execution. If we add up all the timework units of all the project activities, the result is 141 workdays.

The project contains some activities that are performed in parallel, while others are performed in series. Its final duration is 126 time units. If all the activities are sequentially executed (in series) and there are no time buffers, the planned duration will coincide with the planned timework (141 workdays) because each timework unit of each activity is assigned a weight of 1 unit in the EDM.

In order to assign a given probability distribution function to the project activities, Hammad et al. [75] conducted a comparative study on different probability distribution functions (PDFs) and concluded that the most appropriate PDF for the project was normal distribution. However, we decided to use the triangular distribution function for modeling the activity's duration because the definition of the project activities provides us with the most probable, most optimistic, and most pessimistic values. Furthermore, the literature mentions that a triangular distribution can be used as a proxy for beta distribution in the risk analysis [28,76].

Uncertainty about the duration of activities is modeled here as a triangular distribution function whose parameters are the pessimistic, most likely, and optimistic duration of each activity.

Fig. 5 represents the data of both the planned project (TPD) and the real project underway with the curves TAD and TED (as explained in Fig. 1). It can be seen that they are almost linear.

In the project planning phase, we use the Monte Carlo simulation to obtain *N* simulated projects according to the uncertainty of each activity duration. In this example, we employ the commercial Matlab software to generate 25 000 different simulated projects. The software application assigns a random duration to each project activity according to its probability distribution function in each simulation. If we repeat this operation 25 000 times, we will obtain the same number of different projects generated by the Monte Carlo simulation, each with a different final duration. The set of all these projects simulated at their final instant can be represented as a point cloud (see the set of blue dots in Fig. 6).

---

† For a detailed study of these algorithms, see Refs. [78–82].

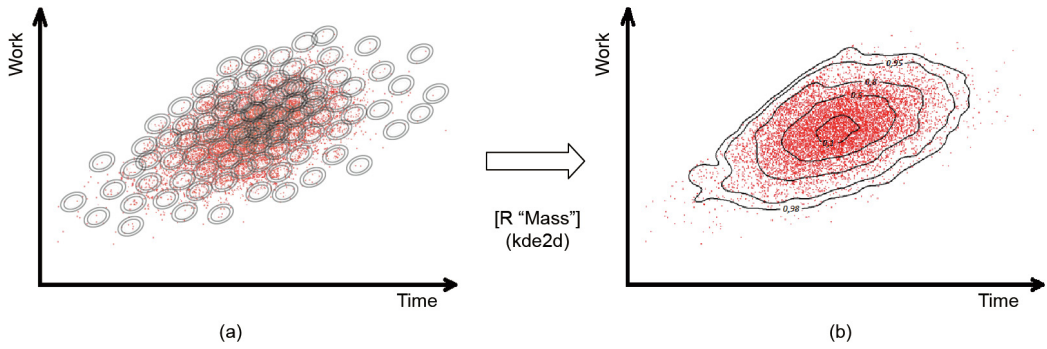

**Fig. 3.** Kernel density estimation. (a) The point cloud (AD$_j$, TAD$_{AD_j}$); (b) probability density curves.

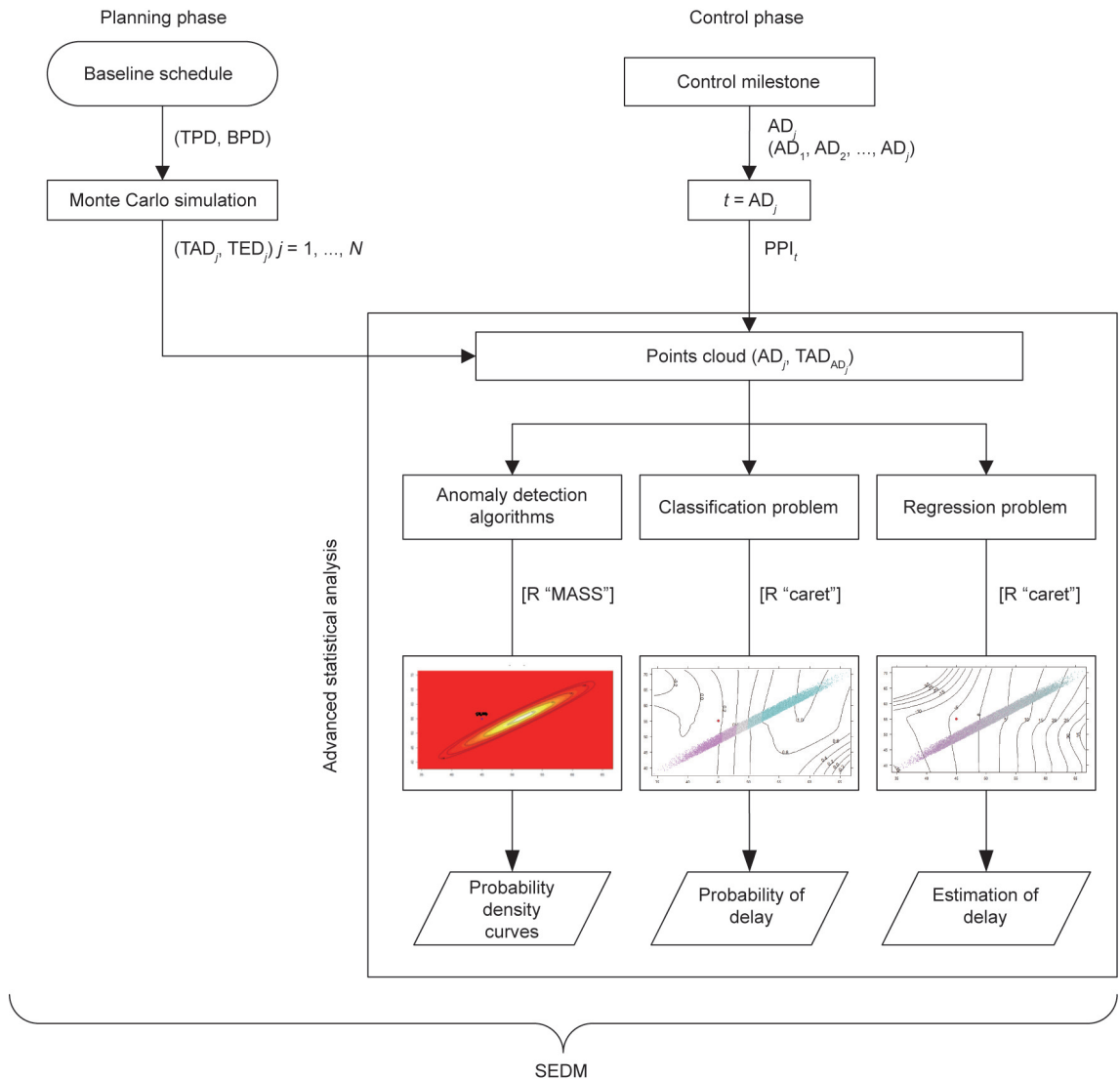

**Fig. 4.** Flow diagram: SEDM.

### 4.1. Deviations in the control milestones

We monitor the deviations in the control milestones as AD = 45 d. We take the data corresponding to the timework units spent by each activity until that control period. At this time, TPD is 48 workdays, TAD is 55 workdays, and TED is 52.54 workdays. We apply Eq. (1) and determine the value of ED($t$) to be 49.54 days $(ED(t) = 45 + (52.54 − 52)/(53 − 52) = 49.54)$. We then apply Eq. (2) and determine the PPI value to be 39.32% (PPI = 49.54/126 = 39.32%). In the planning phase and after applying the Monte Carlo simulation, we obtain 25 000 simulated projects (based on the planned project data).

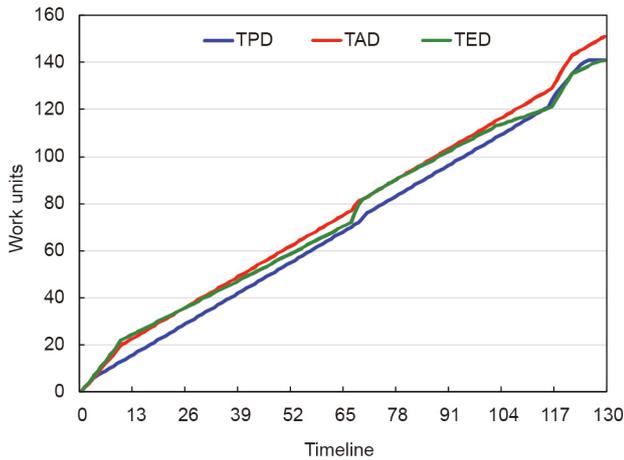

**Fig. 5.** Representation of the project run according to the EDM.

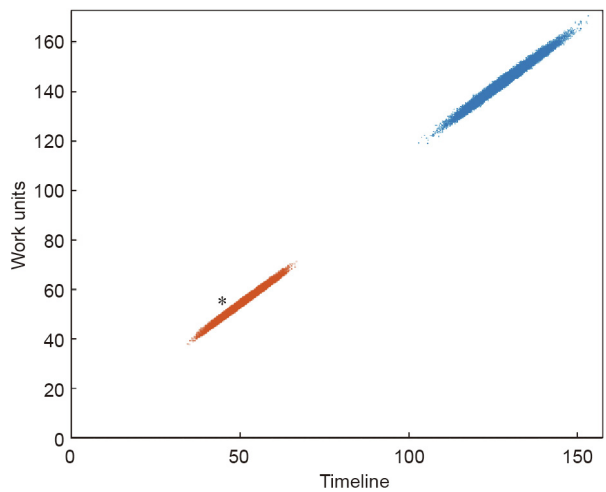

**Fig. 6.** Representation of the simulated projects at both the control time (PPI = 39.32%) and final execution time (PPI = 100%).

In the execution phase, the real project is at a certain control time (AD = 45 d and PPI = 39.32%). For each $j$ simulated project $(P_j)$ and for this control time (PPI = 39.32%), we must calculate which pairs of points $(AD_j, TAD_{AD_j})$ fulfil this condition.

For this purpose, we calculate the value of $TED_{AD}$ of the real project underway. This value will coincide with the same parameter of all the simulated projects $(TED_{AD} = TED_{AD_j})$ (Fig. 2). With the obtained $TED_{AD_j}$ value, for each project $P_j$ we calculate the corresponding time value $AD_j$ and its real duration value $TAD_{AD_j}$. These two values are included in Table 1 for each simulated project.

The columns in Table 1 (AFD and TAD) are the final results of the duration and work periods for each simulated project $(P_j)$. To compare these results with the planned values, we assign a value of 0 if the project is not behind schedule, and a value of 1 otherwise. In the last two columns, we include the actual value of the simulated project's advance or delay $(P_j)$ in relation to the planned one.

For example, in the simulated project identified as $P_1$, PPI = 39.32% corresponds to a pair of values $(AD_j = 53.522,$

$TAD_{AD_j} = 58.107)$. The AFD is 125.135 days, which is earlier (0.865 days) than initially planned (this means that delay = 0).

We can represent the pairs $(AD_j, TAD_{AD_j})$ of the universe of simulated projects for PPI = 39.32% (Fig. 6), which produces the orange-colored cloud of dots at the bottom left of the figure. An asterisk ($^*$) denotes the position in the diagram of the project underway for this PPI value: (AD = 45, $TAD_{AD}$ = 55). If the actual project underway is in the point cloud of the simulated projects (the red-dotted cloud in Fig. 6), then the real project is underway, according to the normal variability of the activities composing the project. If the actual project (represented by $^*$ in Fig. 6) is located outside the point cloud, then this situation cannot be explained by the aleatoric uncertainty of the activities. This situation's execution deviates from the normal variability provided by the set of project activity uncertainties. The blue-colored cloud of dots in the top right of the Fig. 6 represents the final time $(AFD_j, TAD_j)$ for the universe of simulated projects.

We can acquire further information (Fig. 7) about the real project underway (PPI = 39.32%) if we represent the orange-colored cloud of dots $(AD_j, TAD_{AD_j})$ in the control milestone (AD = 45, $TAD_{AD}$ = 55) among the simulated projects that end in advance (delay = 0, $AFD_j \leq BPD$), and among those that end after a delay (delay = 1, $AFD_j > BPD$). The project underway comes closer to the projects that finish with a delay and over workload.

We apply the anomaly detection algorithm (the kde2d function in R "MASS") to the dataset of couples $(AD_j, TAD_{AD_j})$ that corresponds to AD = 45 (PPI = 32.39%) in order to gain more detailed information about the deviation observed in the project that is underway. This function provides the kernel density estimation that is displayed in Fig. 8. We observe that the project underway performs worse than 98% of the simulated projects, which can be taken as a warning to take corrective actions.

### 4.2. Project estimations

To provide some insight into when the project will end, we propose studying the project as both a classification problem and a regression problem. We use the R "caret" package to provide estimations of the probability of the project ending on time (i.e., not being delayed) and the expected eventual final time. Before obtaining the results, we process data by dividing the total sample (25 000 simulation results) into an explicit training dataset used to prepare the model (80% of outcomes) and an unseen test dataset to evaluate the model's performance on unseen data (20% of the results).

Many different metrics can be used to evaluate machine learning algorithms in R. When "caret" is used to evaluate the models, we get distinct metrics as output for classification problems and different metrics for regression problems. The default metrics used are "Accuracy" for classification problems and root mean squared error (RMSE) for regression.

Solving the classification problem, Fig. 9 shows the results returned by R "caret" for our dataset of simulated projects in AD = 45 (PPI = 32.39%) with the already proposed algorithms. The Fig. 9 represents two different ways of measuring the accuracy of the applied algorithms: Accuracy and Kappa. Accuracy and Kappa are the default metrics used to evaluate algorithms in binary and multiclass classification datasets in "caret." Accuracy is the percentage of correctly classified instances among all the instances. Kappa or Cohen's Kappa is similar to a classification accuracy, except that it is normalized at the baseline of the random chance on the dataset.

The bottom of Fig. 9 shows the two types of employed metrics placed in two different quadrants: Accuracy on the left of the figure and Kappa on the right. The different evaluated proposed algorithms are placed on the ordinate axis. The order of placement is descending, according to the accuracy of each algorithm. In the graph, it is possible to compare the averages and see the overlap of the margins between algorithms. These are useful plots because they show both the mean estimated accuracy and the 95% confidence interval (i.e., the range within which 95% of the observed scores fall).

**Table 1**
A sample of the obtained Monte Carlo simulation results and PPI = 39.32%.

| $P_j$ | Simulation results | | | | BPD = 126 | | TPD = 141 | |
|---|---|---|---|---|---|---|---|---|
| | Control time | | Finished project | | Yes (1) / No (0) | | Quantification | |
| | $AD_j$ | $TAD_j$ | AFD | TAD | Delay | Overwork | Delay | Overwork |
| $P_1$ | 53.522 | 58.107 | 125.135 | 141.748 | 0 | 1 | −0.865 | 0.748 |
| $P_2$ | 46.995 | 51.986 | 125.177 | 142.343 | 0 | 1 | −0.823 | 1.343 |
| $P_3$ | 50.529 | 53.453 | 125.053 | 140.014 | 0 | 0 | −0.947 | −0.986 |
| $P_4$ | 51.961 | 55.968 | 132.052 | 148.910 | 1 | 1 | 6.052 | 7.910 |
| $P_5$ | 44.213 | 49.246 | 121.713 | 139.252 | 0 | 0 | −4.287 | −1.748 |
| $P_6$ | 50.653 | 53.861 | 126.977 | 142.506 | 1 | 1 | 0.977 | 1.506 |

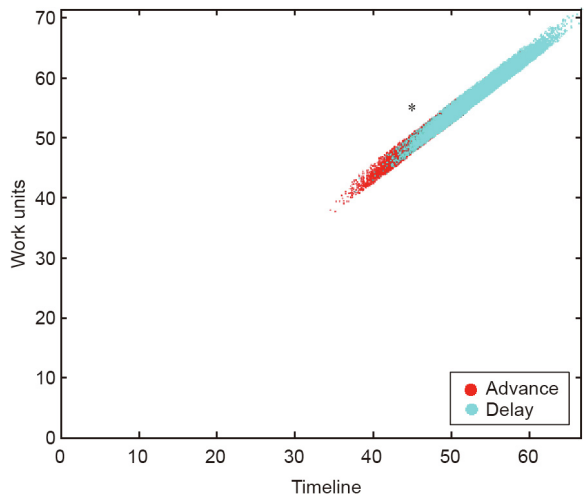

**Fig. 7.** Projects simulated in AD = 45 (PPI = 39.32%). Red dots denote those that will end in advance; cyan dots depict those that will end after a delay.

It can be seen that SVM with a radial basis function is the best option for this case because its average accuracy is the highest (0.80115) at the 0.95 confidence level. R "caret" also provides the same results in a tabular format, including the accuracy achieved by each algorithm (Fig. 10).

This SVM algorithm provides a 38.35% probability at AD = 45 that the project underway will finish after a delay. In other words, there is a 61.65% probability of the project ending early. The regression problem returns an estimation of the expected time when the project will end. Once again, we use R "caret" to select the best algorithm for our problem, and we acquire (in this case) information as a figure (Fig. 11) or as a numerical table (Fig. 12).

RMSE and $R^2$ are the default metrics used to evaluate algorithms in the regression datasets in "caret." The RMSE is the average deviation of the predictions from the observations; it is useful for gaining a general idea of how well (or not) an algorithm is doing in the output variable units. $R^2$, which is known as R-squared or called the coefficient of determination, provides a goodness-of-fit measure for the predictions to observations. This value lies between 0 and 1 for no fit and a perfect fit, respectively. The RMSE provides a general idea of how wrong all the predictions are (where 0 is perfect), and $R^2$ indicates how well the model fits the data (where 1 is perfect and 0 is not well). Fig. 11 represents the error obtained by each regression algorithm applied in our project, sorted in ascending order as the error increases.

We select glmnet as the best algorithm (mean absolute error (MAE) and RMSE are lower). When we apply the glmnet algorithm to our dataset at AD = 45, we obtain an expected delay of −0.577 days.

### 4.3. Validation

The analysis of the project underway with SEDM at AD = 45 can be summarized as follows: There is a 0.6165 probability that the project will end early, the expected time of the final duration is 0.577 days before the BPD (130 days), and the project goes beyond

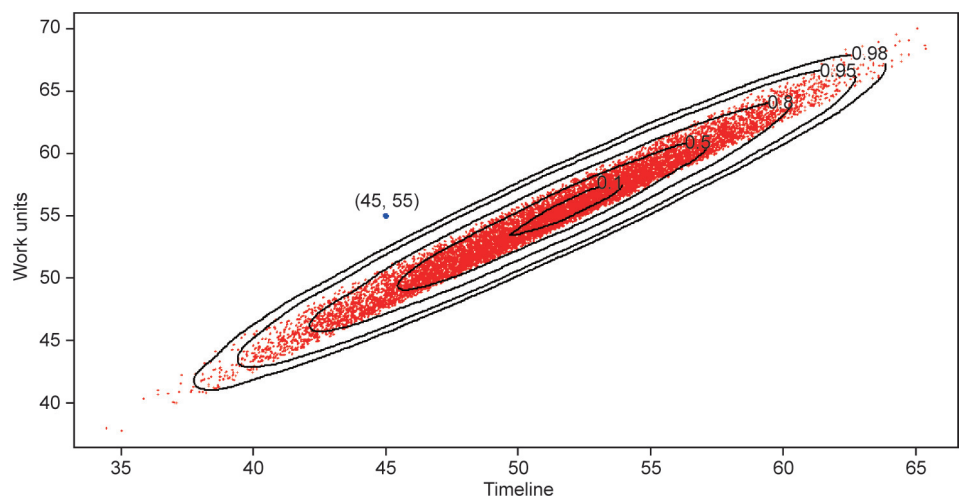

**Fig. 8.** Probability density curves and the project underway.

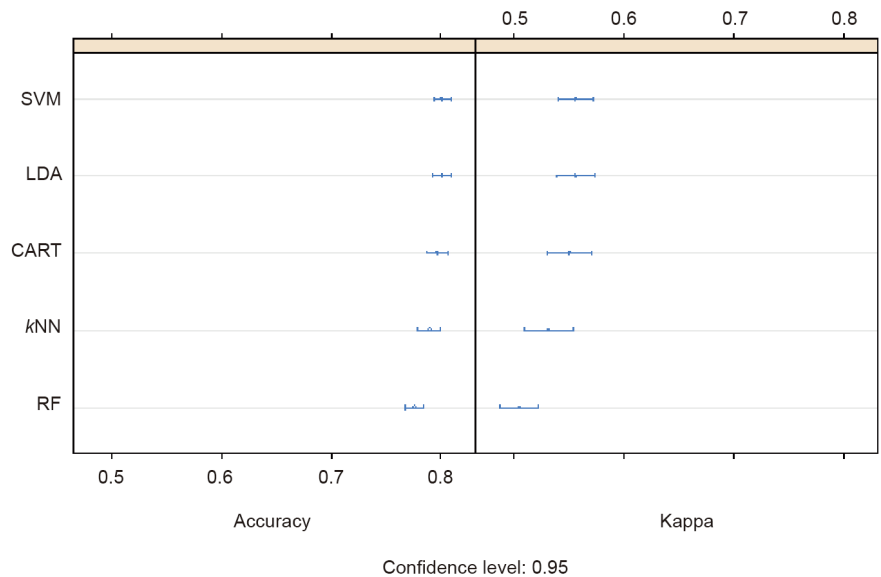

**Fig. 9.** Classification problem: selecting the best algorithm for the case study.

Call:

Summary. resamples (object = results)

Models: LDA, CART, *k*NN, SVM, RF

Number of resamples: 10

Accuracy

|      | Min | 1st Qu | Median | Mean | 3rd Qu | Max | NA's |
|------|------|------|------|------|------|------|------|
| LDA  | 0.7900000 | 0.7936318 | 0.7988741 | 0.8010302 | 0.8022495 | 0.8285357 | 0 |
| CART | 0.7815231 | 0.7869409 | 0.7938819 | 0.7961583 | 0.7999374 | 0.8260325 | 0 |
| *k*NN | 0.7640449 | 0.7813560 | 0.7913826 | 0.7894056 | 0.7951919 | 0.8122653 | 0 |
| SVM  | 0.7912500 | 0.7965044 | 0.7976272 | 0.8011555 | 0.8037500 | 0.8272841 | 0 |
| RF   | 0.7615481 | 0.7657723 | 0.7768750 | 0.7761544 | 0.7837500 | 0.7940075 | 0 |

Kappa

|      | Min | 1st Qu | Median | Mean | 3rd Qu | Max | NA's |
|------|------|------|------|------|------|------|------|
| LDA  | 0.5320780 | 0.5380816 | 0.5521878 | 0.5563190 | 0.5603760 | 0.6169475 | 0 |
| CART | 0.5074130 | 0.5342324 | 0.5481871 | 0.5506419 | 0.5670223 | 0.6088038 | 0 |
| *k*NN | 0.4748853 | 0.5180715 | 0.5347738 | 0.5315967 | 0.5459177 | 0.5823023 | 0 |
| SVM  | 0.5304108 | 0.5451010 | 0.5512033 | 0.5567600 | 0.5635054 | 0.6100178 | 0 |
| RF   | 0.4719933 | 0.4846698 | 0.5063023 | 0.5047193 | 0.5159040 | 0.5474189 | 0 |

**Fig. 10.** Classification problem: data in a tabular format. Min: minimum; Qu: quartile; Max: maximum; NA: not available.

the expected behavior (98% of the universe of simulated projects). This is the analysis that is done only at this specific control milestone.

To validate our proposal, we performed a comparison with two other methodologies, ESM [9] and stochastic EV methodology (SEVM) [15], throughout the time that the project lasts. Table 2 shows the final project duration predictions obtained by the ESM and SEVM methods, as well as the work periods when applying the proposed SEDM method. All these values were calculated at a certain percentage of the project execution.

Fig. 13 shows the project's total duration estimation as computed daily from the beginning to the end, AD = {1, 2, …, 130}, by the three methodologies (the real project ended on day 130). At the control milestone AD = 45, the three methodologies forecast that the project will end early (before the 130 days of this project's AD).

Upon reviewing the data of the real project that is underway, it can be seen that the project is progressing better than planned, until the final periods. This observation is consistent with both the SEDM and ESM forecasting, although ESM is too optimistic. Both SEVM and SEDM forecast quite well. At times, however, SEVM forecasts that the project will end early and, at other times, that it will end with a delay.

We previously adopted MAPE (Eq. (3)) to compare the accuracy of the estimation made by different methodologies. Fig. 14 illustrates this value during the project's life cycle. ESM performs worse than SEVM and SEDM. For this case study, ESDM and ESVM offer similar estimations, with an average error of around 5%.

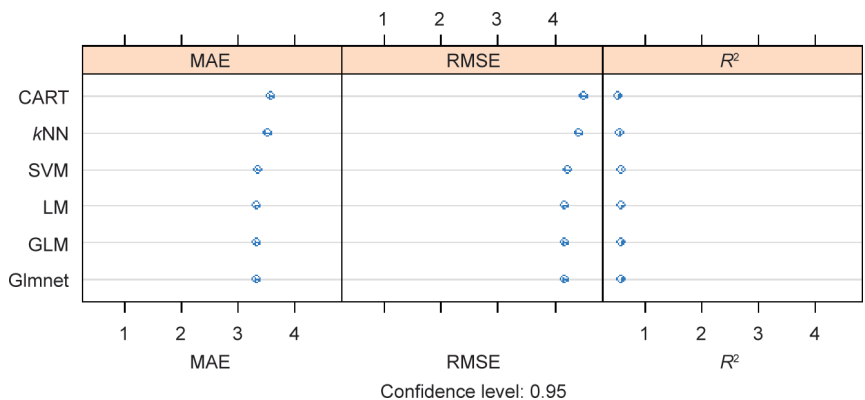

**Fig. 11.** Classification problem: selecting the best algorithm for the case study. MAE: mean absolute error.

> Summary (results)

Call:

Summary. resamples (object = results)

Models: LM, GLM, Glmnet, SVM, CART, $k$NN
Number of resamples: 30

MAE

|      | Min      | 1st Qu   | Median   | Mean     | 3rd Qu   | Max      | NA's |
|------|----------|----------|----------|----------|----------|----------|------|
| LM   | 3.195724 | 3.254230 | 3.320354 | 3.342360 | 3.418672 | 3.543396 | 0    |
| GLM  | 3.195724 | 3.254230 | 3.320354 | 3.342360 | 3.418672 | 3.543396 | 0    |
| Glmnet | 3.194670 | 3.255740 | 3.318696 | 3.341716 | 3.418217 | 3.541175 | 0    |
| SVM  | 3.200200 | 3.288974 | 3.349472 | 3.374492 | 3.461550 | 3.560669 | 0    |
| CART | 3.430582 | 3.515760 | 3.568076 | 3.598175 | 3.667093 | 3.839889 | 0    |
| $k$NN | 3.309923 | 3.473381 | 3.533513 | 3.531273 | 3.620875 | 3.730225 | 0    |

RMSE

|      | Min      | 1st Qu   | Median   | Mean     | 3rd Qu   | Max      | NA's |
|------|----------|----------|----------|----------|----------|----------|------|
| LM   | 3.944744 | 4.118357 | 4.175682 | 4.188504 | 4.239396 | 4.452619 | 0    |
| GLM  | 3.944744 | 4.118357 | 4.175682 | 4.188504 | 4.239396 | 4.452619 | 0    |
| Glmnet | 3.943413 | 4.117564 | 4.174741 | 4.188202 | 4.238181 | 4.452744 | 0    |
| SVM  | 3.951211 | 4.167227 | 4.217994 | 4.237794 | 4.325444 | 4.481123 | 0    |
| CART | 4.270703 | 4.398771 | 4.476339 | 4.498624 | 4.592983 | 4.841628 | 0    |
| $k$NN | 4.155697 | 4.373963 | 4.430295 | 4.429387 | 4.491791 | 4.678946 | 0    |

$R^2$

|      | Min       | 1st Qu    | Median    | Mean      | 3rd Qu    | Max       | NA's |
|------|-----------|-----------|-----------|-----------|-----------|-----------|------|
| LM   | 0.5774464 | 0.5903301 | 0.6039751 | 0.6069097 | 0.6236759 | 0.6430149 | 0    |
| GLM  | 0.5774464 | 0.5903301 | 0.6039751 | 0.6069097 | 0.6236759 | 0.6430149 | 0    |
| Glmnet | 0.5777368 | 0.5903114 | 0.6041077 | 0.6069851 | 0.6236913 | 0.6429746 | 0    |
| SVM  | 0.5649480 | 0.5825936 | 0.5962248 | 0.5978703 | 0.6143537 | 0.6360936 | 0    |
| CART | 0.4995255 | 0.5280143 | 0.5497875 | 0.5465516 | 0.5677085 | 0.5782095 | 0    |
| $k$NN | 0.5243237 | 0.5454060 | 0.5647150 | 0.5636183 | 0.5794646 | 0.6066811 | 0    |

**Fig. 12.** Regression problem: data in a tabular format.

**Table 2**
Estimated final project duration based on ESM, SEVM, and SEDM.

| Control time | AD | ESM | SEVM | AD | SEDM |
|---|---|---|---|---|---|
| 0 | 0 | 126.00 | 126.00 | 0 | 126.00 |
| 10% | 18 | 87.34 | 140.30 | 7 | 122.37 |
| 20% | 41 | 112.30 | 125.19 | 17 | 122.32 |
| 30% | 52 | 117.57 | 130.67 | 31 | 123.27 |
| 40% | 63 | 121.26 | 127.06 | 45 | 125.42 |
| 50% | 68 | 119.90 | 125.98 | 60 | 126.29 |
| 60% | 77 | 115.50 | 122.91 | 69 | 124.32 |
| 70% | 86 | 117.32 | 124.03 | 81 | 123.14 |
| 80% | 94 | 118.77 | 124.33 | 95 | 123.80 |
| 90% | 105 | 121.22 | 124.64 | 112 | 126.93 |
| 100% | 130 | 130.00 | 130.00 | 130 | 130.00 |
| PPI = 39.32% | 45 | 114.22 | 127.18 | 45 | 125.42 |

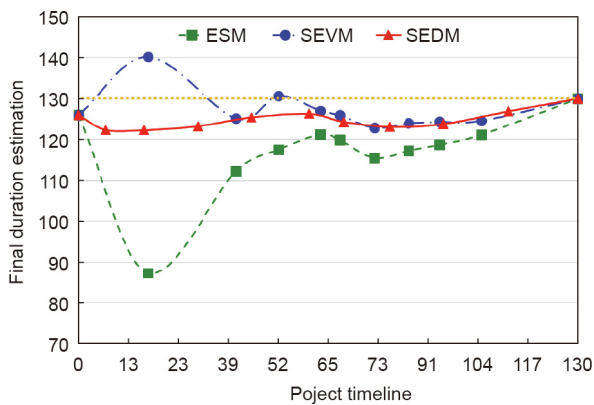

**Fig. 13.** The final project duration estimations during the project's timeline.

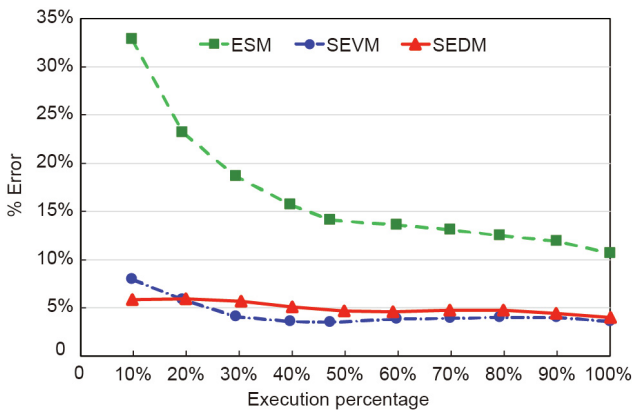

**Fig. 14.** MAPE: the final project duration estimation.

## 5. The project control and monitoring benchmark: SEDM versus SEVM and ESM

SEDM is a methodology used for monitoring and controlling projects. In the previously presented case study, both SEDM and SEVM provided similar final project duration estimations. To compare the two, we review the accuracy of the estimations (MAPE) in real project samples, which we select from among the projects available in the OR-AS database [72–74]. The database contains baseline scheduling data (network, resources, etc.), risk analysis data (for Monte Carlo simulations), and project control data (using EVM and ES metrics).

We select four projects with different topology networks (serial/parallel (SP) indicator), TPD, and number of activities: 2012–10, 2014–08, 2016–15, and 2016–28. The SP indicator has already been used [13,29,70,74,77], and its definition is shown in Eq. (4):

$$SP = \frac{n_s - 1}{n_t - 1} \tag{4}$$

where $n_s$ is the number of serial paths and $n_t$ is the total number of paths (including parallel paths). The SP value ranges from 1 to 0, where SP = 0 refers to a 100% parallel project network, while SP = 1 represents a 100% serial project network.

Table 3 shows the data for each project: the identifier, SP indicator ($s/p$), $n_s$, $n_t$, number of critical paths ($N$), BPD, AFD, TPD, and TAD.

We also include the MAPE diagram for the whole project life cycle for ESM, SEVM, and SEDM (Fig. 15). The first conclusion is that stochastic methods (i.e., SEDM and SEVM) give better estimations than ESM when managing stochastic projects. In comparison, both SEDM and SEVM exhibit similar behavior for the 2012–10 and 2016–15 projects. These projects take different values for BPD, AFD, TPD, TAD, and number of paths, but the SP indicator is over 0.5 (i.e., it is a serial project network). The 2014–08 and 2016–28 projects have similar SP indicators (0.41 vs 0.45). SEVM performs better than SEDM for the 2014–08 project, but SEDM performs better than SEVM for the 2016–28 project. Thus, the first result of this comparison is that it makes sense to use SEDM once it is that both methodologies return different estimations.

These results lead to a new research question: Can we establish a rule to assess whether the SEDM or the SEVM should be used to project monitoring and control? What factors affect each method's accuracy?

**Table 3**
Data for different project networks.

| Project | $s/p$ | $n_t$ | $N$ | $n_s$ | BPD | AFD | TPD | TAD |
|---|---|---|---|---|---|---|---|---|
| 2012–10 | 0.823 | 18 | 4 | 15 | 54 | 60 | 59 | 67 |
| 2014–08 | 0.410 | 40 | 37 | 17 | 233 | 275 | 402 | 496 |
| 2016–15 | 0.666 | 13 | 12 | 9 | 126 | 130 | 141 | 151 |
| 2016–28 | 0.450 | 21 | 6 | 10 | 71 | 76 | 151 | 161 |

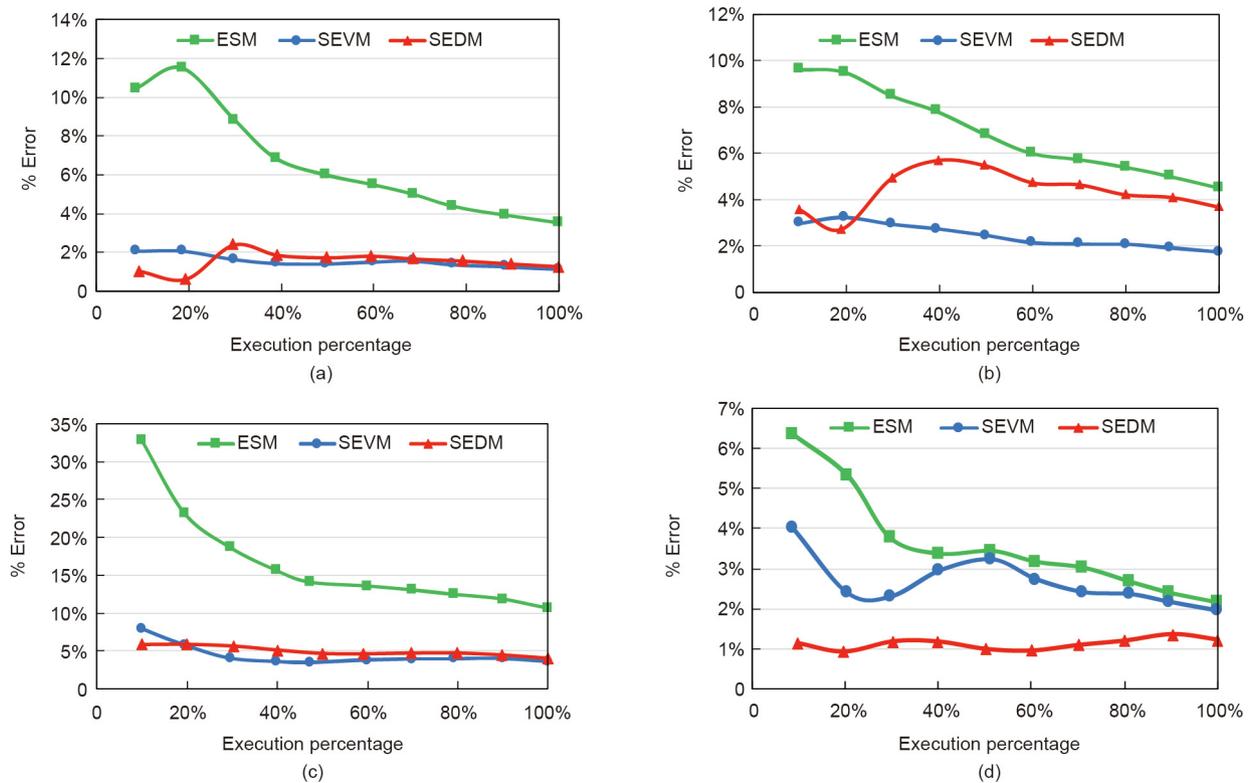

**Fig. 15.** The MAPE study for different project networks. (a) 2012–10; (b) 2014–08; (c) 2016–15; (d) 2016–28.

## 6. Conclusions

The EDM facilitates project monitoring and control when the cost of activities is not a good estimator for the project duration. As an alternative to ESM and to SEVM, which use costs, SEDM is based on workload. SEDM is useful for monitoring and controlling projects in which costs are not the key indicator, such as development cooperation projects, and projects for which it is difficult to make a clear costs estimation: large and complex infrastructure projects, or research and development projects. In such cases, uncertainty in estimating the cost of activities comes into play, and using workload and SEDM can provide relevant information for project managers.

Some studies explain how to implement EDM in deterministic projects, or even in fuzzy projects. However, situations can arise in which—due to the stochastic nature of the project activities' duration—activities become aleatory. Therefore, the time required for an entire project to finish is always questionable. Nevertheless, no method exists in the literature to apply EDM to stochastic projects. By meeting this need, the present research makes it possible to introduce uncertainty into project activities and to be able to use EDM methodology for the control of stochastic projects. This methodology allows project managers to know the project status at each control time (delay/advance). It also allows them to determine whether project overruns are within the expected variability or if there are structural and systemic changes throughout the project life cycle. Finally, this approach allows anomalous situations regarding the project definition to be detected by considering the possible correlation between the time and cost of project activities. It also makes it possible to calculate the probability of exceeding the expected duration.

In this paper, we explained how SEDM should be applied to stochastic projects in a similar way to SEVM in order to monitor and make good final project duration estimations. Monte Carlo simulation provides better project duration estimations than ESM when working with stochastic projects. The process can be implemented in real projects, allowing project managers to monitor relevant deviations in workload or duration estimations.

We presented a case study to illustrate the application of SEDM to stochastic projects, and validated the SEDM methodology in the proposed case study. We compared the results of SEDM with the results of both ESM and SEVM. Finally, we explained our analysis on the interest and usability of SEDM. We selected a set of four projects for which SEVM offers good final project duration estimations. We also found that SEDM and SEVM sometimes return equivalent final duration estimations, although one occasionally performs better than the other.

Further research is required to determine how the network topology (SP indicator) and other parameters affect the accuracy of both SEDM and SEVM. While a high SP indicator value (close to 1) can imply that SEDM and SEVM are equivalent and other parameters do not affect the accuracy of either SEDM or SEVM, other project parameters are necessary for SP values under 0.5.

## Acknowledgments

This research has been partially financed by the Regional Government of Castille and Leon (Spain) with Grant (VA180P20).

## Compliance with ethics guidelines

Fernando Acebes, David Poza, José Manuel González-Varona, and Adolfo López-Paredes declare that they have no conflict of interest or financial conflicts to disclose.